\documentclass[prb,preprint]{revtex4}

\usepackage{graphicx}% Include figure files \usepackage[draft]{graphicx} for quick compilation- no figures
\usepackage{amsfonts,amssymb, amsmath}  %amsmath: \eqref{} outputs eq number in blackets () ;-)
\usepackage{dcolumn}% Align table columns on decimal point
\usepackage{bm}% bold math
\usepackage{caption}

\usepackage{subfig}
\begin{document}

\title{Comparison of EIT schemes in semiconductor quantum dot structures: \\ Impact of many-body interactions}

%\author{J. Houmark}
%\email{jakob.houmark@nanotech.dtu.dk}
%\affiliation{DTU Nanotech - Department of Micro- and Nanotechnology, Technical University of Denmark, \O rsteds Plads, DK-2800 Kongens Lyngby, Denmark}
%
%\author{T. R. Nielsen}
%\affiliation{DTU Fotonik - Department of Photonics Engineering, Technical University of Denmark, \O rsteds Plads, DK-2800 Kongens Lyngby, Denmark}

\author{J. Houmark$^{1,}$\footnote{Electronic address: jakob.houmark@nanotech.dtu.dk}, T. R. Nielsen$^2$,  J. M{\o}rk$^2$ and A.-P. Jauho$^{1,3}$}

\address{$^1$ DTU Nanotech - Department of Micro- and Nanotechnology, Technical University of Denmark, \O rsteds Plads, DK-2800 Kongens Lyngby, Denmark}
\address{$^2$ DTU Fotonik - Department of Photonics Engineering, Technical University of Denmark, \O rsteds Plads, DK-2800 Kongens Lyngby, Denmark}
\address{$^3$ Department of Applied Physics, Helsinki University of Technology, P. O. Box 1100, 02015 HUT, Finland}

\begin{abstract}
   We investigate the impact of many-body interactions on group-velocity slowdown achieved via
   Electromagnetically Induced Transparency (EIT) in quantum dots using three
   different coupling-probe schemes (Ladder, $V$ and $\Lambda$ , respectively).
   We find that for all schemes many-body interactions have an important impact on the slow light properties.
   In the case of the $\Lambda$ and V schemes, the minimum required coupling power to achieve slow light is significantly
   reduced by many-body interactions. $V$ type schemes are found to be generally preferable, due to a favorable redistribution
   of carriers in energy space.
\end{abstract}
\maketitle

\section{Introduction}
Quantum dot (QD) based materials are promising for applications
exploiting quantum coherence phenomena due to their atom-like
properties and long dephasing times.\cite{art:Borri_dephasing_times}
They have been proposed to act as active media in devices for controlling the emission pattern of phased array
antennas\cite{art:Mork_laser_photon_2008} or in slow light based all--optical
buffers.\cite{art:hasnain_optical_buffer}

A particular physical effect that can be utilized for generating slow light is
electromagnetically induced transparency (EIT). EIT refers to an artificially created spectral region of transparency
in the middle of an absorption line due to the destructive
quantum interference arising from two transitions
in a three-level system.\cite{art:Harris_prl_1990, art:Harris_phys_today_1997} By virtue of the Kramers-Kr\"{o}nig relations such an absorption reduction is accompanied by a large positive slope of the refractive index which translates into a reduced group velocity in vicinity of the resonance.

Very recently, the first experimental studies of EIT in QD systems have been performed \cite{art:Saulius_apl_2008} where a coherent absorption dip in a coupling-probe experiment has been observed for an optically thin structure.
Semiconductor QD based EIT schemes without real
carrier excitations have been studied using models from atomic physics.\cite{art:hasnain_optical_buffer, art:Kim_slolight_condmat, art:Janes_2008, art:Kær_opt_express} Such EIT
configurations involve pumping of intraband transitions whose
wavelengths lie in the deep infrared, a regime for which high
intensity laser operation is very difficult. Carrier--exciting schemes using interband coupling transitions
therefore come into play. Recently, a theoretical description with
the inclusion of many-body effects for a solid state QD EIT $\Lambda$ configuration has been reported.\cite{art:michael_apl, art:chow_jmo}

Using carrier-exciting schemes one in effect addresses two types of quantum coherence phenomena,
EIT as well as coherent population oscillation\cite{art:CPO_Bigelow_science_2003,art:Chuang_prb_2005} (CPO).
CPO is a four-wave mixing effect based on interference between the coupling and probe fields.
One should therefor keep in mind that schemes involving carrier excitation
generally would contain a mixture of the two effects. However, CPO can be ruled out by choosing a setup utilizing orthogonal
polarization directions for the coupling and probe field thus preventing the possibility of interference.

Concerning the EIT effect; an inherent problem of the carrier--exciting scheme is
that the carriers excited by the coupling field block the transitions via the
Pauli blocking factor, effectively decreasing the strength of the
transitions, making such configurations less attractive than
those that do not excite carriers. In addition, the excited carriers
modify the spectral properties of the system via their mutual
Coulomb interaction. Such effects cannot be accounted for in the
non-interacting model. The work presented in Refs.~\onlinecite{art:michael_apl} and \onlinecite{art:chow_jmo}
addresses the many-body aspects of a carrier exciting $\Lambda$ configuration in a transient regime.
The study of pulse propagation in a semiconductor slow light medium would generally involve solving the coupled Maxwell-Bloch equations. However, under certain circumstances an analysis of the steady state properties of the semiconductor Bloch equations alone is adequate. In the sense that the linear optical response extracted in this limit is directly linked to the propagation characteristics of a wavepacket traveling in an optically thick QD system. Non carrier--exciting schemes have only been studied with the inclusion of many-body effects in this limit on one occasion, \cite{art:Houmark_jpc_2008} while studies of carrier--exciting schemes have not, to the best of our knowledge, been discussed in the literature.
The aim of this paper is to present a
comparison between different EIT schemes, with and without carrier
excitation, that can be realized in the same dot structure. We study
the EIT generated slow light properties of InAs QDs by
solving the generalized semiconductor Bloch equations (SBE) in the
Hartree-Fock approximation. The slowdown capabilities of the Ladder,
$V$ and $\Lambda$ schemes (see Fig.~\ref{3_level_schemes})
obtained in steady state are compared using two models; the atomic
model where interactions are disregarded, and the interacting model
where many-body effects are taken into account.

\section{Theoretical model}
\begin{figure}[here]
   \begin{center}
   \includegraphics[width=0.7\textwidth]{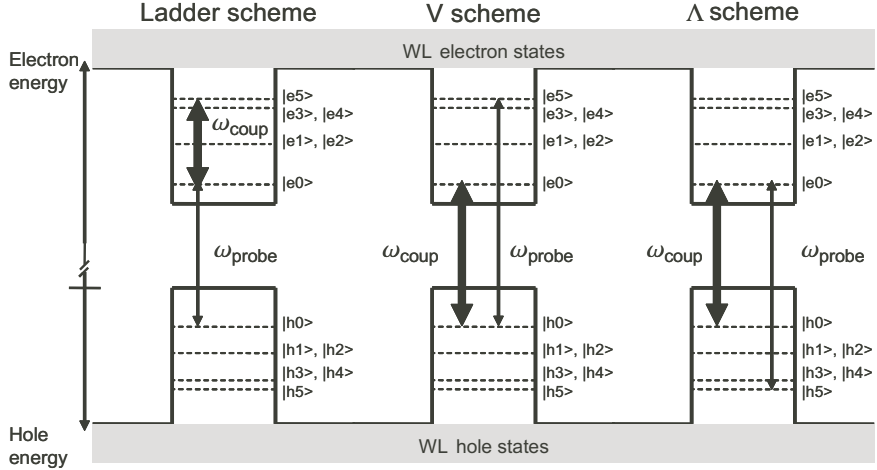}\\
   \end{center}
   \caption{Schematic quantum dot level structure and three EIT configurations. The frequency of the intense coupling field is denoted $\omega_{coup}$ while the weak probe field is shown as $\omega_{probe}$. For illustrative purposes the figure has not been drawn to scale.} \label{3_level_schemes}
\end{figure}

The heterostructure under consideration consists of conical InAs
quantum dots (radius of 9 nm and height 3 nm) residing on a 1.2 nm
thick wetting layer (WL), sandwiched between two slabs of GaAs.
The electronic structure is calculated as the solution to the
single--band Schr\"{o}dinger equation for the envelope wavefunction
in the effective mass approximation.\cite{art:Melnik_QD_nanotech_2004} Using
effective electron and hole masses $m_e = 0.067 ~m_0$, $m_h = 0.15 ~m_0$
and a conduction/valence band offset $\mathrm{CBO} = 705 ~\mathrm{meV}$, $\mathrm{VBO} = 363 ~\mathrm{meV}$, we
find six confined hole states (labeled $|\mathrm{h}0\rangle$ to $
|\mathrm{h}5\rangle$) as well as six confined electron states
(labeled $|\mathrm{e}0\rangle $ to $ |\mathrm{e}5\rangle$), all
doubly degenerate due to spin. Furthermore, the inherent rotational
symmetry ensures complete degeneracy of the first and second
excited- as well as third and fourth excited state for both bands.
For each band we also find the onset of a continuous set of delocalized
states extending into the wetting layer. These WL states are treated as plane waves.
The resulting energy level structure along with three different
EIT-schemes are shown in Fig.~\ref{3_level_schemes}.
The dipole selection rules allow for the realization of the three
archetypical EIT schemes; Ladder, $V$ and $\Lambda$. In the two
latter cases, the coupling field excites an interband
transition, resulting in optical pumping of the dot.
We model experiments where a continuous wave coupling field is
irradiating a homogeneous ensemble of QD's.
The steady state system response is obtained by applying a
weak probe pulse with a gaussian envelope at times later than any other
timescale of the system relative to the onset of the coupling field,
such that transient effects may be neglected. In order to effectively utilize the slowdown of the light,
as in an all--optical buffer, the probe field must propagate
within the plane of the active medium. Assuming the QD's to lie in the $x$-$y$ plane,
we take the probe polarization along the $x$-axis
and the propagation direction along the $y$-axis. To completely rule out CPO effects
we need orthogonal polarizations of the coupling and probe fields.
%The probe pulse is in
%all cases both polarized along and propagating in the QD plane ($x$
%and $y$ directions, respectively).
For the the $V$ and $\Lambda$ schemes the coupling field is polarized in the $y$-direction, but in
these situations we let the coupling field propagate perpendicular to the QD
plane so that we can disregard propagation effects, e.g.~attenuation, in the coupling field.
In the Ladder scheme the coupling field connects states $|\mathrm{e}0\rangle$ and
$|\mathrm{e}5\rangle$ and is polarized in the growth direction of
the QD's ($z$-direction) and thus traveling in the QD plane. We disregard
propagation effects, as the coupling field is effectively
connecting two empty states, thus rendering the transition transparent.

The interband dipole moments connecting $|\mathrm{e}0\rangle$ and $|\mathrm{h}5\rangle$ as well as
$|\mathrm{e}5\rangle$ and $|\mathrm{h}0\rangle$ are non-zero even in the one-band effective mass description.
A detailed strain based 8-band ${\bf k} \cdot {\bf p}$ calculation\cite{art:JHN_DB_kdotp_dipolemoments}
shows however that these dipole moments are roughly a factor of 20 larger compared to the one-band result.
As our focus is on the influence of the many-body interactions on the slow down effects we will therefore assume the
${\bf k} \cdot {\bf p}$ based results for the $\mu_{e0h5} = 10.24~e\mathrm{\AA} $ and $\mu_{e5h0} = 10.14~e\mathrm{\AA}$ dipole moments.
The other relevant dipole moments are $\mu_{e0h0} = 15.55~ e\mathrm{\AA}$ and $\mu_{e0e5} = 2.79~ e\mathrm{\AA}$.

%In order to compare the different EIT
%configurations, the strength of the interband transitions connecting
%$|\mathrm{e}0\rangle$ and $|\mathrm{h}5\rangle$ as well as
%$|\mathrm{e}5\rangle$ and $|\mathrm{h}0\rangle$ have been increased
%from $\mu_{e0h5} = 0.51 ~e\mathrm{\AA}$ to $\mu_{e0h5} = 10.24~e\mathrm{\AA} $
%and $ \mu_{e5h0} = 0.50 ~e\mathrm{\AA}$ to $\mu_{e5h0} = 10.14~e\mathrm{\AA}$ , respectively.
%This ad hoc procedure is justified by
%calculations using ${\bf k} \cdot {\bf p}$ theory showing an
%increase of similar magnitude for these particular transitions
%relative to the ground state interband transition $(\mu_{e0h0} = 15.55~ e\mathrm{\AA})$,
%when using a more accurate 8 band model that also incorporates shear
%strain\cite{art:JHN_DB_kdotp_dipolemoments}.

The linear optical response to the probe, i.e.\ the susceptibility
$\chi(\omega)$, whose real and imaginary part are related to
refraction and absorption, respectively, is found from the
macroscopic polarization $P(\omega)$ as $\chi(\omega)  =
\frac{P(\omega)}{\epsilon_0 E_\mathrm{p}(\omega)}$, where
$\epsilon_0$ is the vacuum permittivity and $E_\mathrm{p}(\omega)$
is the amplitude of the probe field.

The time resolved macroscopic polarization component in the direction of the
probe field, $P(t)$, is computed from the microscopic polarizations
according to semiclassical theory:\cite{book:Haug_Koch}
\begin{eqnarray}
    P(t) & = & \frac{1}{w}\Big(N_\mathrm{dot} \sum_{i,j} \mu_{ij} P_{ij}(t)
    + \frac{1}{A}\sum_\mathbf{k} \left[\mu_\mathbf{k} P_\mathbf{k}(t) + c.c. \right)] \Big) .
\end{eqnarray}
Where $P_{ij}$ and $P_{k}$ are microscopic polarization components of
localized dot states ($i, j$) and diagonal interband polarization components of
delocalized WL states ($\mathbf{k}$), respectively. In this treatment we
disregard polarization components relating to transitions connecting dot and
WL states. Dipole matrix elements between localized
states are denoted $\mu_{ij}$, whereas $\mu_\mathbf{k}$ is the dipole moment
relating to WL states. $N_\mathrm{dot}$ is the two-dimensional
density of the dots in the WL plane, $A$ is the normalization area
of the WL, and $w$ is the thickness of the active region.

The microscopic polarizations are the off-diagonal components
$\Psi_{\nu_1 \nu_2}$ ($\nu_1 \neq\nu_2$) of the reduced density matrix $\rho_{\nu_1 \nu_2}$, where $\nu$ refers to either a QD state $i$ or a WL state $\mathbf{k}$. The time development of the polarizations are found by solving the SBE in the Hartree-Fock approximation, see e.g.\ Ref.~\onlinecite{thesis:TRN}, given
(in the electron-electron picture for the sake of brevity) by
\begin{eqnarray}\label{eq:bloch}
   i \hbar \frac{\partial}{\partial t}\Psi_{\nu_1 \nu_2}(t) &-& \left[ \tilde{\epsilon}_{\nu_1}(t) - \tilde{\epsilon}_{\nu_2}(t) \right]\Psi_{\nu_1 \nu_2}(t) - \left[ n_{\nu_2}(t) - n_{\nu_1}(t) \right]\Omega_{\nu_1 \nu_2}(t) \nonumber \\
   & - & \sum_{\nu_3 \neq \nu_1, \nu_2} \left[ \Omega_{\nu_1 \nu_3}(t) \Psi_{\nu_3 \nu_2}(t) -  \Psi_{\nu_1 \nu_3}(t) \Omega_{\nu_3 \nu_2}(t) \right] \nonumber \label{eq_dephasing}\\
   &=& i \hbar S_{\nu_1 \nu_2}(t) \approx - i \hbar \gamma_d \Psi_{\nu_1 \nu_2}, \\
   i \hbar \frac{\partial}{\partial t} n_{\nu_1}(t) &-& \sum_{\nu_3 \neq \nu_1} \left[ \Omega_{\nu_1 \nu_3 }(t) \Psi_{\nu_3 \nu_1}(t) - \Omega_{\nu_3 \nu_1 }(t) \Psi_{\nu_1 \nu_3}(t) \right] \nonumber \\
   & = & i \hbar S_{\nu_1 \nu_1}(t) \nonumber \\
   & \approx & - i \hbar \gamma_{nr} n_{\nu_1}(t) - i \hbar \gamma_{c-c} [ n_{\nu_1}(t) - f_{\nu_1}(\mu_p, T_p) ] - i \hbar \gamma_{c-p} [ n_{\nu_1}(t) - f_{\nu_1}(\mu_l, T_l) ]\, ,
\end{eqnarray}
where
\begin{eqnarray}
    \tilde{\epsilon}_\nu(t)  &=&  \epsilon_\nu + \sum_{\nu_3 \nu_4} \left[ V_{\nu \nu_4 \nu_3 \nu} - V_{\nu \nu_4 \nu \nu_3} \right] \rho_{\nu_3 \nu_4}(t) \\
    \Omega_{\nu_1 \nu_2}(t) &=& - e \mu_{\nu_1 \nu_2}E(t) + \sum_{\nu_3 \nu_4}\left[ V_{\nu_1 \nu_4 \nu_3 \nu_2} - V_{\nu_1 \nu_4 \nu_2 \nu_3} \right] \rho_{\nu_3 \nu_4}(t) ,
\end{eqnarray}
are the Hartree-Fock renormalized single particle energy and
generalized Rabi frequency, respectively.
$n_\nu$ is the diagonal component of the density matrix, i.e.\ $\rho_{\nu \nu}$.
The term $- e \mu_{\nu_1 \nu_2}E(t)$ is the electromagnetic field interaction in
the dipole approximation, and the matrix elements of the Coulomb interaction are
$V_{\nu_1\nu_2\nu_3\nu_4} = \int  \Phi^{\ast}_{\nu_1}(\mathbf{r}) \Phi^{\ast}_{\nu_2} (\mathbf{r}') e^2/\epsilon_0 \epsilon_b | \mathbf{r} -\mathbf{r}'|  \Phi_{\nu_3}(\mathbf{r}') \Phi_{\nu_4}(\mathbf{r}) \, d^3 r d^3 r' $. Coulomb elements are found by approximating the numerically
evaluated localized dot states by those of a harmonic oscillator; and $V_{\nu_1\nu_2\nu_3\nu_4}$ is then calculated following Refs.~\onlinecite{art:TRN_prb_2004} and \onlinecite{art:TRN_prb_2005}.
For the situations considered here, screening effects are disregarded due to low WL densities.

Off-diagonal scattering terms $S_{\nu_1 \nu_2}(t)$ are approximated
by a temperature dependent effective dephasing rate $\gamma_d$;
\begin{eqnarray}
    S_{\nu_1 \nu_2}(t) &  \approx &  - \gamma_d \Psi_{\nu_1 \nu_2} \, .
\end{eqnarray}
Diagonal terms representing collision induced particle exchange processes,
are mimicked by a nonradiative recombination and a population relaxation
towards quasi-equilibrium Fermi-Dirac functions $f_\nu$~\cite{art:Mork_opt_soc_1996}, determined by the charge-carrier density and temperature. The scattering rates are
denoted $\gamma_{c-c}$ and $\gamma_{c-p}$ representing carrier-carrier
and carrier-phonon scattering, and the recombination rate is called $\gamma_{nr}$;
\begin{eqnarray}
    S_{\nu_1 \nu_1}(t)  &\approx &  - \gamma_{nr} n_{\nu_1}(t) -\gamma_{c-c}[ n_{\nu_1}(t) - f_{\nu_1}(\mu_p, T_p) ] \nonumber \\
      & & - \gamma_{c-p}  [ n_{\nu_1}(t) - f_{\nu_1}(\mu_l, T_l) ]\,
      .
\end{eqnarray}
Here $\mu$ and $T$ are the chemical potential and temperature of
either the plasma ($p$) or lattice ($l$), which are found following the procedure presented in Ref.~\onlinecite{art:chow_pra}. We arrive at the non-interacting
(atomic) model by taking the limit where all Coulomb
elements and population scattering rates are set to zero.
The results presented here use a dot density of $N_{\mathrm{dot}} = 5 \cdot 10^{14} \; \mathrm{m}^{-2}$, a discretization of the WL into 100 $k$-points
and a fixed lattice temperature of 200 K, for which the literature~\cite{art:Borri_deph_phys_rev_B,art:Borri_dephasing_times}
gives scattering rates around $\gamma_d= 1.5\cdot 10^{12} \; \mathrm{s}^{-1}$, $\gamma_{c-c}= 2.0\cdot 10^{12} \; \mathrm{s}^{-1}$, $\gamma_{c-p}= 2.0\cdot 10^{11} \;\mathrm{s}^{-1}$, and $\gamma_{nr}= 1.0\cdot 10^{9} \;\mathrm{s}^{-1}$.

\begin{figure}
        \begin{center}
        \subfloat[{Ladder scheme.} ]
        {
        \label{chi_ladder_ss}
        \includegraphics[width=0.45\textwidth]{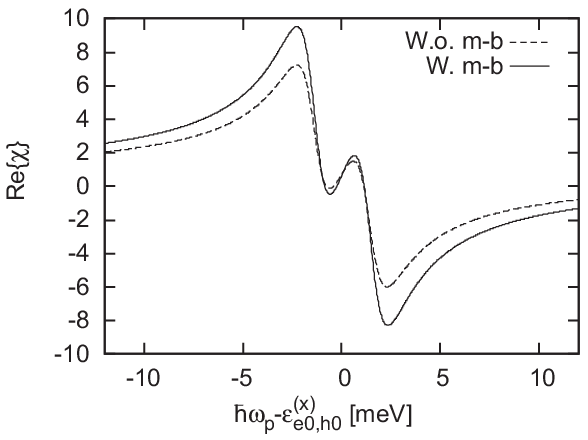}
        %\subfloat[Imaginary part of the susceptibility.]
        \includegraphics[width=0.45\textwidth]{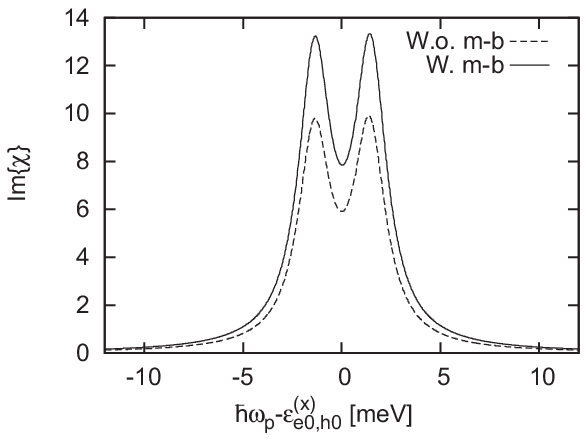}
        }\\
        \subfloat[{$V$ scheme.}]
        {
        \label{chi_vee_ss}
        \includegraphics[width=0.45\textwidth]{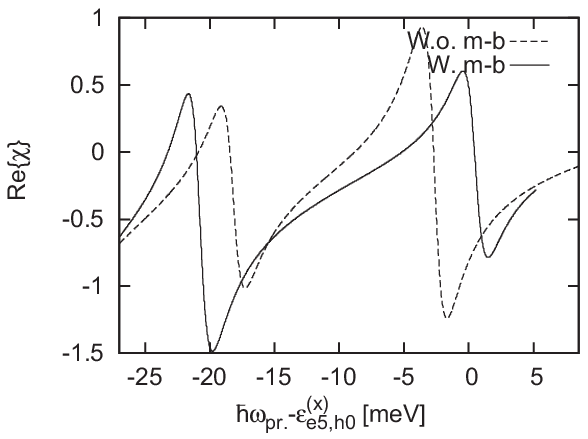}
        %\subfloat[Imaginary part of the susceptibility.]
        \includegraphics[width=0.45\textwidth]{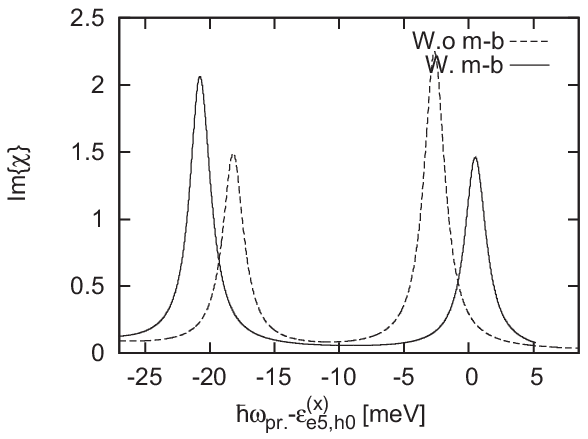}
        }\\
        \subfloat[{$\Lambda$ scheme.}]
        %\subfloat[Real part of the susceptibility.]
        {
        \label{chi_lambda_ss}
        \includegraphics[width=0.45\textwidth]{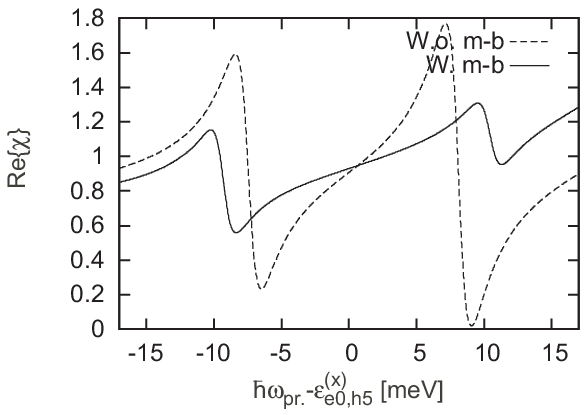}
        %\subfloat[Imaginary part of the susceptibility.]
        \includegraphics[width=0.45\textwidth]{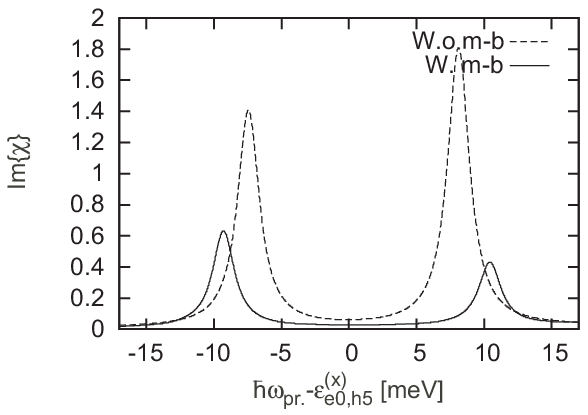}
        }
        \end{center}
    \caption{Ladder, $V$ and $\Lambda$ schemes using a coupling field
    intensity of $26 \; \mathrm{MW}/\mathrm{cm}^2$: Complex susceptibility vs. probe energy. The dashed line corresponds to the atomic model, i.e.,
    \underline{without} many-body interactions, while the solid line is evaluated \underline{with} many-body interactions. The energy is measured relative to the zero density excitonic resonances $\epsilon^{(x)}_{e0,h0}$, $\epsilon^{(x)}_{e5,h0}$ and $\epsilon^{(x)}_{e0,h5}$, respectively. To facilitate a comparison between the two cases, the noninteracting spectra have been shifted accordingly.}\label{fig:chi_3_schemes}
\end{figure}
\section{Results}
The optical response for the three different schemes using a coupling field
with an intensity of $26 \; \mathrm{MW}/\mathrm{cm}^2$ is shown in
Fig.~\ref{fig:chi_3_schemes} illustrating both the interacting and the
non-interacting cases. An immediate difference between the atomic and many-body approach is the change in probe
field energy towards negative detuning. This is due to the excitonic
shift of the various probe transitions.
An apparent feature of the Ladder scheme (Fig.~\ref{chi_ladder_ss}) is that the peaks of the
imaginary part of the susceptibility are highest for the many-body model.
While the distance between the peaks remains the same one can readily see
that a larger area is covered by the many-body spectrum. This is evidence
that oscillator strength has been shifted into the $|e0\rangle-|h0\rangle$
transition, in other words it has been Coulomb enhanced. Considering the
the real part of the susceptibility, the enhancement of the probe
transition results in more pronounced features, notably a larger slope
at zero detuning. Both curves are generally shifted upwards from the zero point;
this behavior is due to transitions in the vicinity of the probe which
are contributing to the background index of the area in question.

%can again be attributed to a change in the dot oscillator strengths, which tends to take "weight" from high energy transitions and put into lower energy transitions. The wetting layer states surrounding the dot plays non-negligible part in this redistribution. The upwards shift happens because the transitions in the vicinity of the probe transition are generally enlarged, thereby increasing their contribution to the "background index" of the area in question.

The optical responses for $V$ and $\Lambda$ schemes are shown in
Fig.~\ref{chi_vee_ss} and Fig.~\ref{chi_lambda_ss}, respectively.
For the interacting case the coupling field has been tuned to the
zero density exciton resonance of the probe transition.
%\begin{figure}[h]
%    \centering
%        %\subfloat[Real part of the susceptibility.]
%        {
%        \includegraphics[width=0.45\textwidth]{figs/chi_re_vee_ss_1E7_opt_express.eps}
%        }
%        %\subfloat[Imaginary part of the susceptibility.]
%        {
%        \includegraphics[width=0.45\textwidth]{figs/chi_im_vee_ss_1E7_opt_express.eps}
%        }
%    \caption{V scheme: Complex susceptibility vs. probe energy. The dashed line retains to the atomic model, the solid line to the many-body model. The energy is measured relative to the zero density excitonic resonance $\epsilon^{(x)}_{e5,h0}$. To facilitate a comparison between the two cases, the noninteracting spectra have been shifted.} \label{chi_vee_ss}
%\end{figure}
%
%
%\begin{figure}[h]
%    \centering
%        %\subfloat[Real part of the susceptibility.]
%        {
%        \includegraphics[width=0.45\textwidth]{figs/chi_re_lambda_ss_1E7_opt_express.eps}
%        }
%        %\subfloat[Imaginary part of the susceptibility.]
%        {
%        \includegraphics[width=0.45\textwidth]{figs/chi_im_lambda_ss_1E7_opt_express.eps}
%        }
%    \caption{$\Lambda$ scheme: Complex susceptibility vs. probe energy. The dashed line retains to the atomic model, the solid line to the many-body model. The energy is measured relative to the zero density excitonic resonance $\epsilon^{(x)}_{e0,h5}$. To facilitate a comparison between the two cases, the noninteracting spectra have been shifted.}
%\end{figure}
The asymmetry in the peak heights of the imaginary part of the
susceptibility has different origin for the interacting and noninteracting cases. The skewness
in the atomic model is due to the fact that we are not dealing with
a closed three level system. The control field pumping the
$|e0\rangle-|h0\rangle$ is also connecting the dipole allowed
$|e5\rangle-|h5\rangle$ transition, however severely negatively
detuned. Effectively we are dealing with two EIT schemes, the
original V ($\Lambda$) and a detuned $\Lambda$ (V) scheme. In
general a negatively detuned $\Lambda$ or $V$ EIT scheme has a
prominent shift in peak height towards positive detunings
(resembling an optical Stark shift). What is seen in the two figures
is an admixture of the symmetric peaks owing to the resonant $V$
($\Lambda$) scheme and a Stark shifted transition. Further evidence
of this effect has been obtained by altogether disallowing the
"conflicting" transition in which case one recovers the symmetric
result. In the models including interactions the same asymmetry
should be expected, but it is countered by a negative shift in
resonance energy induced by the Coulomb interaction with the excited
carriers. This means that the coupling is detuned positively with respect to the
resonance, and hence the asymmetry tends towards negative probe
energy. The probe transitions in these cases are not enhanced rather
they are suppressed by the inclusion of many-body effects; this
can be seen by the fact that the features are generally smaller in magnitude than
in the atomic model. The splitting of the peaks is larger
though, showing that the effective Rabi frequency is higher, owing
to the Coulomb enhancement of the $|e0\rangle-|h0\rangle$ coupling
transition.

As a basis for comparing the slow light capabilities of the
different schemes we examine the maximum obtainable slowdown factor
$S$, which is equal to the group index and is a measure of the group velocity reduction. The slowdown factor $S$ is a
figure of merit relevant for optical storage, and is defined via
\begin{eqnarray}\label{slowdown_factor}
    S = \frac{c_0}{v_g} = n + \omega \frac{\partial n }{\partial \omega}\; ,
\end{eqnarray}
where $c_0$ is the speed of light in vacuum, and $n = {\rm Re}  \{
[n_b^2 + \chi(\omega) ]^{\frac{1}{2}} \}$ is the refractive index. The
maximum slowdown is found at the frequency for which the slope of the refractive index is
largest. Notice, that the slowdown factor obtained away from resonance
is given by the background refractive index.
To make a just comparison, we detune the coupling field used
in the many-body $V$ and $\Lambda$ models from the zero density
exciton resonance, so that the peaks of ${\rm Im} \{\chi(\omega)\}$
become symmetric. However, the amount of detuning for this to be realized depends
on the intensity of the coupling field. Changing the intensity in turn
changes the amount of carriers being excited and thus the excitonic shift, which results in different
detunings for different intensities.
\begin{figure}[here]
   \begin{center}
   \includegraphics[width=0.45\textwidth]{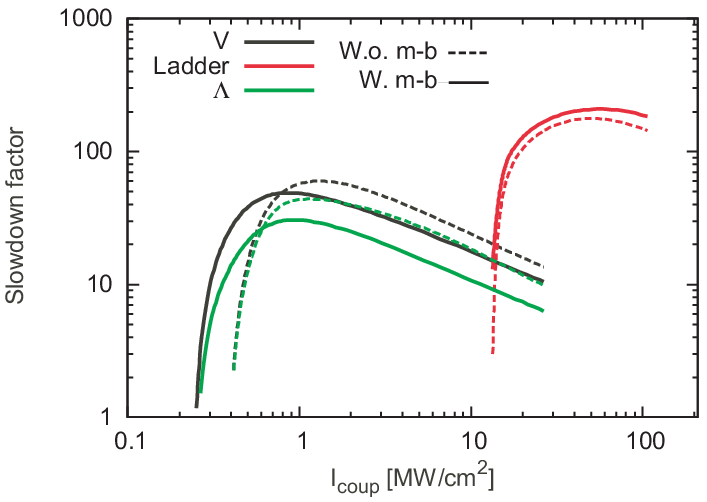}
   \includegraphics[width=0.46\textwidth]{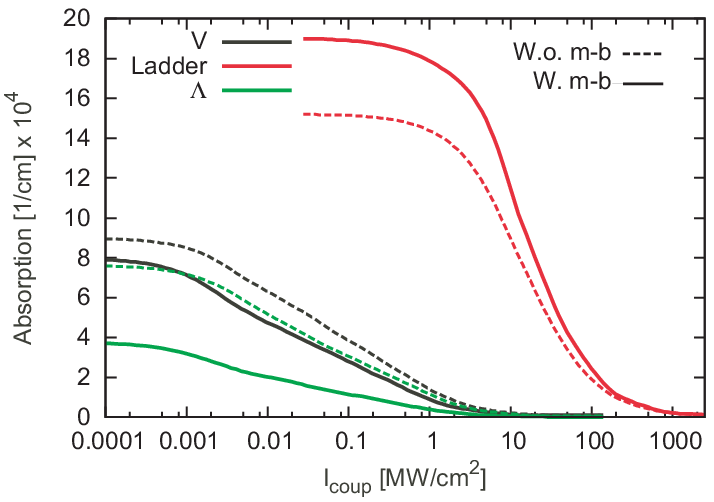}
   \end{center}
   \caption{(Color online) Maximum slowdown factor and corresponding absorption coefficient vs.\ coupling intensity for various EIT-schemes. The dashed curves are for the independent particle (atomic) model, while the solid curves include many-body interactions.} \label{slowdown}
\end{figure}

Figure~\ref{slowdown} depicts the maximum achievable slowdown factor
and the corresponding absorption coefficient $\alpha(\omega) = 2\frac{\omega}{c_0}\mathrm{Im}\{n(\omega)\}$
as a function of coupling power for all three schemes. A striking
feature of Fig.~\ref{slowdown} is that the inclusion of
many-body effects leads to different results depending on the
choice of EIT scheme. The results of the carrier-exciting $V$ and
$\Lambda$ setups are seen to differ fundamentally from the Ladder
scheme. Inspecting the absorption coefficient plot we see two plateaus,
corresponding to the maximum (minimum) absorption in the absence (presence) of EIT.
The transition from the upper to the lower plateau happens across fewer orders of magnitude in the coupling power
for the ladder scheme than the other two schemes. Here the absorption coefficient drop
is solely due to the quantum coherence effect setting in, whereas the transition
for $\Lambda$ and $V$ happens across a significantly larger relative range. At low coupling power
the absorption drop is driven by the excitation of carriers occupying the probe transition.
As evidenced by the slowdown plot, the EIT effect sets
in at larger coupling powers, only when we are near the lower plateau.
The largest slowdown values are achieved using the Ladder
scheme, for which the slowdown factor is increased significantly
when interactions are included. This is due to Coulomb enhancement
of the $|e0\rangle - |h0\rangle$ resonance probed in this scheme.
The slowdown effect is seen to disappear at the same value of
coupling power for both cases, which indicates that the coupling
transition ($|e0\rangle - |e5\rangle$) utilized in the Ladder
configuration is unchanged by the inclusion of many-body
effects. On contrary, for the carrier-exciting schemes V and
$\Lambda$, many-body effects have a significant impact on the coupling
threshold.

Both $V$ and $\Lambda$ schemes show largest slowdown values for the
non-interacting model. On the other hand, the noninteracting model
overestimates the minimum required coupling power for observing slow light
by roughly a factor of two compared to the more realistic case of interacting particles.
As both schemes utilize the same coupling transition they experience
the same coupling power threshold, in both the interacting as well
as the non-interacting case. The shift in required coupling power
can be attributed solely to the Rabi energy enhancement of the
coupling transition. This conclusion is reached by inspecting the
absorption spectrum in absence of a coupling field. By comparing the
height of the $|e0\rangle - |h0\rangle$ resonance (coupling
transition used in both schemes) with and without interactions we
find that the dipole moment of the transition is enlarged by roughly
a factor of $1.2$. If we, in the atomic model, enlarge the coupling
dipole moment by the same amount, we end up with a result having the
same minimal requirement on coupling power as the interacting case.
This result stands in contrast to the findings in Ref.~\onlinecite{art:michael_apl},
where a shift in required coupling power,
due to Coulomb enhancement, of two orders of magnitude was reported. However, this work
was performed in a transient regime and a direct comparison is therefore not applicable here.

\begin{figure}[here]
    \begin{center}
    \includegraphics[width=0.5\textwidth]{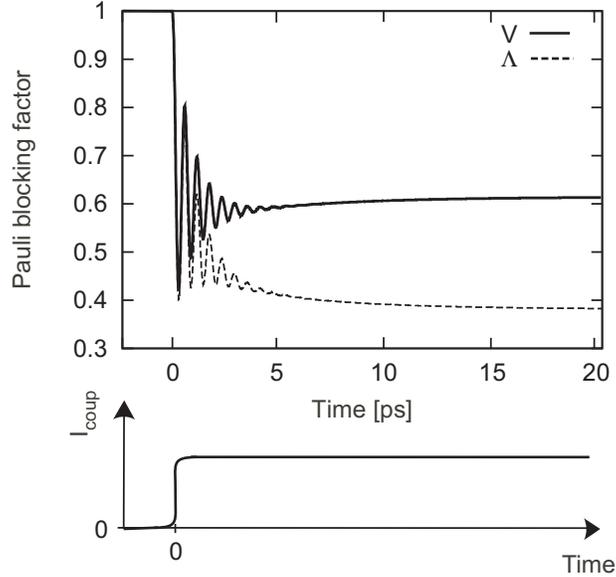}\\
    \end{center}
    \caption{Temporal development of the Pauli probe blocking factors for $\Lambda$ and $V$ schemes at coupling intensity $2.5 \; \mathrm{MW}/\mathrm{cm}^2$.
    Shown in the figure are the Rabi oscillations of the populations at the onset of the coupling field and the relaxation towards their stationary value. Also shown is an illustration of the "turn on" of the coupling laser.} \label{Pauli_block_ss}
\end{figure}

The $V$ scheme is preferable to the $\Lambda$ scheme, due to its
higher slowdown values. The reason is twofold. Firstly, based on
observations from the absorption spectrum without coupling, we find
that the two probe transitions are both Coulomb suppressed, however
the $\Lambda$ scheme to a higher degree than $V$. Secondly the fact
that the $V$ probe connects a hole ground state to an electron excited
state results in a larger Pauli blocking factor (the third term in
equation (\ref{eq:bloch})) as compared to the $\Lambda$ scheme.
Figure~\ref{Pauli_block_ss} demonstrates this for the interacting
case utilizing a coupling intensity of $2.5 \;
\mathrm{MW}/\mathrm{cm}^2$.
For the interacting model the redistribution of carriers plays a
crucial role; the smaller effective electron mass leads to a larger
energy spacing of the electronic levels, which means that the
electron excited states become less populated than their hole
counterparts. In the case presented here the hole ground state is
seen to be depleted and carriers are redistributed into the higher
lying energy states. For electrons the redistribution is less
prominent, and as the coupling field excites more and more carriers,
electrons accumulate in the ground state. Thus the Pauli blocking
factor seen by the probe in the $V$ configuration is always the
larger, which ultimately translates into an increased slowdown
factor. This result is quite general and could act as a pointer for
experimental realization of EIT mediated slow light.

\section{Conclusion}
In conclusion, we have investigated the slow light properties of
InAs QDs using a model including many-body effects for
three different EIT schemes and found fundamental differences.
%Emphasis has been put on the differences between using an atomic model or one that includes
%many-body interactions on the level of the Hartree-Fock approximation.
The Ladder scheme that utilizes a transparent coupling transition has its
slowdown factor increased due to Coulomb enhancement. However, there
is observed no change in the necessary coupling power required to
reach EIT. Conversely in the $V$ and $\Lambda$ schemes, many-body
effects enhance the coupling transition resulting in a lowering of the
necessary coupling power. The $V$
type configuration is found to be preferable, due to a favorable redistribution of carriers. \\

This work has been supported by the Danish Research Council for
Technology and Innovation through the project QUEST and the European Commission
through the IST project "QPhoton" (Contract No. IST-29283). APJ is grateful to the FiDiPro program of the Finnish Academy during
the final stages of this work.

%\section*{References}
\providecommand{\newblock}{}

%\bibliographystyle{iopart-num}
%\bibliography{quantum-coherence-references}

\end{document}